\documentstyle[epsfig,eqsecnum,aps]{revtex}
\newcommand{\mbbf}[1]{{\mbox{\bf{#1}}}}

\begin{document}
\draft

%%%---- title ----%%%
\title{Determinant of a new fermionic action on a lattice - (II)}
\author{A. Takami, T. Hashimoto, M. Horibe, and A. Hayashi}
\address{Department of Applied Physics, Fukui University, Fukui 910}
\maketitle

%%%%%%% abstract %%%%%%%
\begin{abstract}
We investigate the fermion determinant of a new action
on a $(1+D)$-dimensional lattice for SU(2) gauge groups.
This action possesses the discrete chiral symmetry
and provides $2^D$-component fermions.
We also comment on the numerical results on fermion determinants
in the $(1+D)$-dimensional SU(3) gauge fields.
\end{abstract}

%%%%%%% introduction %%%%%%%
\section{introduction}
As is well known the lattice formulation of fermions
has extra physical particles or breaks the chiral symmetry.
This is unavoidable under a few plausible assumptions \cite{NN}.
Several methods have been proposed to deal with this difficulty.
Wilson's formulation \cite{Wil}, which is one of the most popular schemes,
eliminates the unwanted particles with an additional term which vanishes
in the naive continuum limit.
However, this formulation sacrifices the chiral symmetry.
An alternative scheme is the staggered fermion formulation
proposed by Kogut and Susskind \cite{KS}.
This scheme preserves the discrete chiral symmetry
and in this point the staggered fermion has an advantage
over the Wilson fermion.
But, the staggered formulation describes
a theory with $2^{\frac{1+D}{2}}$ degenerate quark flavours
($2^{1+D}$ components) in $(1+D)$ dimensions,
while there is no restriction on the flavour number
in the Wilson formulation.
Recently, it has been shown that lattice fermionic actions
with the Ginsparg-Wilson relation \cite{GW} have an exact chiral symmetry
and are free from restriction on the flavour number.
But, these actions cannot be "ultralocal" \cite{Ultra-Local},
which makes numerical simulations complicated.

In the recent papers \cite{Our-Formula1,Our-Formula2}, we proposed
a new type of fermionic action on a $(1+D)$-dimensional lattice.
The action is ultralocal and has discrete chiral symmetry.
On the Euclidean lattice the minimal number of fermion components is $2^D$,
which should be compared with $2^{1+D}$ of the staggered fermion.
When dynamical fermions are included, the numerical feasibility
relies on the reality and positivity of the fermion determinant.
In the previous paper \cite{Our-Simulation1} we investigated, analytically
and numerically, the fermion determinant of our new action
in the $(1+1)$-dimensional U(1) lattice gauge theory.
We showed the reality of our fermion determinant under the condition
fixing the global phase of link variables along the temporal direction.
By a similar discussion to the U(1) gauge group,
we could also find the reality and the positivity of our fermion determinant
in the $(1+1)$-dimensional SU(N) lattice gauge theory.

In this paper we analytically show that our fermion determinant
with the SU(2) gauge fields is real and positive in $(1+D)$ dimensions.
We also comment on the numerical results of the fermion determinant
in the $(1+D)$-dimensional SU(3) gauge fields, and discuss
the effectiveness of our new action for SU(2) and SU(3) lattice gauge theories.

%%%%%%% new fermionic action %%%%%%%
\section{new fermionic action}
In the recent paper \cite{Our-Formula2},
we proposed a new fermionic action on the Euclidean lattice.
Though the action keeps the discrete chiral symmetry
like the staggered fermion action, the fermion field
has $2^D$ components in $(1+D)$ dimensions.
In this section we briefly sketch our formalism for later convenience.

The action can be written with a fermion matrix $\Lambda$ as
\begin{eqnarray}
	S_f = \sum_{n, m}
		\psi^\dagger_n \Lambda_{n m} \psi_m,
	\label{eq:action0}
\end{eqnarray}
where the summation is over lattice points and spinor indices, and our fermion
matrix is defined by
\begin{eqnarray}
	\Lambda = 1 - S_0^\dagger U_E.
	\label{eq:fermi_mat0}
\end{eqnarray}
Here $U_E$ is the Euclidean time evolution operator
and $S_\mu$ is the unit shift operator defined by
\begin{eqnarray}
	S_\mu \psi (x^0, x^1, \ldots, x^\mu, \ldots, x^D)
		 = \psi (x^0, x^1, \ldots, x^\mu+1, \ldots, x^D)
	\qquad	(\mu = 0, 1, \ldots, D).
\end{eqnarray}

We require that the propagator has no extra poles
and find that $U_E$ has the form
\begin{eqnarray}
	U_E = 1 - \sum_{i=1}^D \frac{r_E}{2} \left\{ i X_i
		\left( S_i-S_i^\dagger \right) + \left( 1 - Y_i \right)
		\left( S_i - 2 + S_i^\dagger \right) \right\},
	\label{eq:u_e0}
\end{eqnarray}
where $r_E$ is the ratio of the temporal
lattice constant to the spatial one.
The spinor matrices $X$'s and $Y$'s should satisfy the following algebra: 
\begin{eqnarray}
	\left\{
		\begin{array}{ccl}
		\left\{ X_i, X_j \right\}
		& = & \displaystyle{\frac{2}{r_E}} \delta_{ij}, \\
		\left\{ X_i, Y_j \right\}
		& = & 0, \\
		\left\{ Y_i, Y_j \right\}
		& = & 2 \left( \displaystyle{\frac{1}{r_E}} \delta_{ij} + 1 \right),
		\end{array}
	\right.
	\label{eq:algebraXY}
\end{eqnarray}
where $i$ and $j$ run from $1$ to $D$. 
The matrix $2 (\delta_{ij}/r_E+1) $ is positive definite
for any positive $r_E$, therefore $X$'s and $Y$'s can be assumed hermitian,
\begin{eqnarray}
	X_i^\dagger = X_i, \qquad
	Y_i^\dagger = Y_i.
\end{eqnarray}
The matrices $X$'s and $Y$'s can be expressed by the Clifford algebra:
\begin{eqnarray}
	\Gamma_n^\dagger = \Gamma_n, \qquad
	\left\{ \Gamma_n, \Gamma_m \right\} = 2 \delta_{nm}
	\qquad (n, m = 1, \ldots, 2D)
\end{eqnarray}
in several ways.
One is
\begin{eqnarray}
	X_i = \displaystyle{\sqrt{\frac{1}{r_E}}} \Gamma_i, \qquad
	Y_i = \displaystyle{\sum_{j=1}^D} \alpha_{ij} \Gamma_{D+j},
	\label{eq:XYGamma}
\end{eqnarray}
as was used in the previous paper.
Another one is
\begin{eqnarray}
	X_i = \sqrt{\frac{1}{r_E}} \Gamma_i, \qquad
	Y_i = \sqrt{\frac{1}{r_E}} \Gamma_{D+i} + \Gamma_{2D+1},
	\label{eq:XYGamma2}
\end{eqnarray}
where $\Gamma_{2D+1}$ is
\begin{eqnarray}
    \Gamma_{2D+1} = (-i)^D \Gamma_1 \Gamma_2 \cdots \Gamma_{2D}.
\end{eqnarray}
The latter is more convenient than the former for later use.
The dimension of the irreducible representation for $\Gamma$'s is
$2^D$ and accordingly $\psi$ has $2^D$ components.

The interaction of the fermion with gauge fields is
introduced by replacing the unit shift operators by covariant ones:
\begin{eqnarray}
	S_\mu \to S_\mu(x) \equiv U_{x, x+\hat{\mu}} S_\mu,
	\label{eq:replace}
\end{eqnarray}
where $\hat{\mu}$ is the unit vector along the $\mu$'th direction,
and $U_{x, y}$ is a link variable connecting sites $x$ and $y$.

The fermion matrix Eq.(\ref{eq:fermi_mat0}) and the
time evolution operator Eq.(\ref{eq:u_e0}) become
\begin{eqnarray}
	\Lambda (x) = 1 - S_0^\dagger(x) U_E(x),
	\label{eq:fermi_mat1}
\end{eqnarray}
and
\begin{eqnarray}
	U_E(x) = 1 - \sum_{i=1}^D \frac{r_E}{2} \left\{ i X_i
		\left(S_i(x) - S_i^\dagger(x) \right)
		+ \left( 1 - Y_i \right)
		\left( S_i(x) - 2 + S_i^\dagger(x) \right) \right\}.
	\label{eq:u_e1}
\end{eqnarray}

%%%%%%% fermion determinant for SU(2) case %%%%%%%
\section{fermion determinant for SU(2) case}

In this section we analytically study the determinant
of our fermion matrix in SU(2) gauge fields.

First, in the $(1+1)$-dimensional case, the complex conjugation of $U_E(x)$ is
\begin{eqnarray}
	U_E^{\ast}(x) = 1 - \frac{r_E}{2} \left\{ -i X_1^\ast
		\left( S_1^\ast(x) + S_1^{\ast^\dagger}(x) \right)
		+ \left( 1 - Y_1^{\ast} \right)
		\left( S_1^\ast(x) - 2 + S_1^{\ast^\dagger}(x) \right) \right\},
\end{eqnarray}
where we can write
\begin{eqnarray}
	S_1(x) = \alpha_0(x) \mbox{\bf{1}} + i \sum_{i=1}^3 \alpha_i(x) \tau_i,
\end{eqnarray}
since the link variables in $S_1(x)$ are SU(2) gauge group elements.
Here $\alpha_0(x)$ and $\alpha_i(x)$ are real and depend on lattice points
and $\tau_{1,2,3}$ are the Pauli-matrices:
\begin{eqnarray}
	\tau_1 = \left(
		\begin{array}{cc}
			0 & 1 \\
			1 & 0
		\end{array}
	\right) , \qquad
	\tau_2 = \left(
		\begin{array}{cc}
			0 & -i \\
			i & 0
		\end{array}
	\right) , \qquad
	\tau_3 = \left(
		\begin{array}{cc}
			1 & 0 \\
			0 & -1 
		\end{array}
	\right).
\end{eqnarray}
Then we have
\begin{eqnarray}
	S_1(x) \tau_2 = \tau_2 S_1^\ast(x).
\end{eqnarray}
By the same discussion for other unit shift operators
$S_0^\dagger$ and $S_1^\dagger$, we also have
\begin{eqnarray}
	S_0^\dagger(x) \tau_2 = \tau_2 {S_0^\ast}^\dagger(x), \qquad
	S_1^\dagger(x) \tau_2 = \tau_2 {S_1^\ast}^\dagger(x).
\end{eqnarray}
If we can find the matrix $\Gamma$ such that
\begin{eqnarray}
    \left\{
    \begin{array}{rcl}
		X_1^\ast \Gamma & = & - \Gamma X_1, \\
		Y_1^\ast \Gamma & = & \Gamma Y_1,
    \end{array}
    \right.
	\label{eq:XY_1}
\end{eqnarray}
it is easily shown that
\begin{eqnarray}
	\left( \Gamma \otimes \tau_2 \right)
	\left( S_0^\dagger(x) U_E(x) \right)^\ast
	\left( \Gamma \otimes \tau_2 \right) = S_0^\dagger(x) U_E(x).
	\label{eq:t_Gast_t}
\end{eqnarray}
For example, we make the following choice:
\begin{eqnarray}
    X_1 = \sqrt{\frac{1}{r_E}} \tau_1, \qquad
    Y_1 = \sqrt{\frac{1}{r_E}} \tau_2 + \tau_3,
\end{eqnarray}
the matrix $\Gamma$ acting on two components fermi fields defined by
\begin{eqnarray}
    \Gamma = \tau_3
\end{eqnarray}
satisfies Eq.(\ref{eq:XY_1}).
The Eq.(\ref{eq:t_Gast_t}) implies that if $\lambda$ is some eigenvalue
of our fermion matrix $\Lambda(x)$, then $\lambda^\ast$ is an eigenvalue
of $\Lambda^\ast(x)$ and thus also of $\Lambda(x)$.
Therefore eigenvalues of $\Lambda(x)$ are either real or come
in complex conjugate pairs.
From the above discussion we can prove the reality
of our fermion determinant for the SU(2) gauge groups.

Next we show its positivity.
We define
\begin{eqnarray}
	\Gamma^\prime = \left( \Gamma \otimes \tau_2 \right) K,
\end{eqnarray}
where $K$ is complex-conjugation operator.
We find
\begin{eqnarray}
	\Gamma^\prime \Gamma^\prime
		= \left( \tau_3 \otimes \tau_2 \right) K
			\left( \tau_3 \otimes \tau_2 \right) K
		= \left( \tau_3 \otimes \tau_2 \right)
			\left( \tau_3 \otimes (-\tau_2) \right) = -1,
	\label{eq:GG-1}
	\end{eqnarray}
and from the relation Eq.(\ref{eq:t_Gast_t}) we can show
\begin{eqnarray}
	\left[ \Gamma^\prime, \Lambda(x) \right] = 0.
	\label{eq:Gp_comm}
\end{eqnarray}
For a real eigenvalue $\lambda_R$ of $\Lambda(x)$
and the eigenvector $v_R$ for this eigenvalue,
from Eq.(\ref{eq:Gp_comm}) we obtain
\begin{eqnarray}
	\Lambda(x) \Gamma^\prime v_R
		= \Gamma^\prime \Lambda(x) v_R
		= \Gamma^\prime \lambda_R v_R
		= \lambda_R \Gamma^\prime v_R.
\end{eqnarray}
Suppose $\Gamma^\prime v_R = c v_R$, then we find
\begin{eqnarray}
	{\Gamma^\prime}^2 v_R = \Gamma^\prime c v_R 
		= c^\ast \Gamma^\prime v_R = {\left| c \right|}^2 v_R,
\end{eqnarray}
which is inconsistent with Eq.(\ref{eq:GG-1}),
so that $\Gamma^\prime v_R$ is different eigenvector for the same eigenvalue.
Therefore the eigenvalues on real axis are degenerate in pairs
and the determinant of $\Lambda(x)$ is positive.

The above proof of the positivity of our fermion determinant
for the SU(2) group can be expanded to higher dimensions.
We can make a fundamental representation for the Clifford algebra
with $2D$ elements $\Gamma_n$ ($n = 1, \ldots, 2D$)
using direct products of the Pauli-matrices:
\begin{eqnarray}
	\begin{array}{rcl}
	\Gamma_1     & = & \tau_1 \otimes
		\underbrace{\tau_3 \otimes \cdots \otimes \tau_3}_{D-1} \\
	\Gamma_2     & = & \mbbf{1} \otimes \tau_1 \otimes
		\underbrace{\tau_3 \otimes \cdots \otimes \tau_3}_{D-2} \\
	\vdots       &   & \\
	\Gamma_i     & = &
		\underbrace{\mbbf{1} \otimes \cdots \otimes \mbbf{1}}_{i-1} \otimes
		\tau_1 \otimes \underbrace{\tau_3 \otimes \cdots \otimes \tau_3}_{D-i}
		\\
	\vdots       &   & \\
	\Gamma_D     & = &
		\underbrace{\mbbf{1} \otimes \cdots \otimes \mbbf{1}}_{D-1}
		\otimes \tau_1 \\
	\Gamma_{D+1} & = & \tau_2 \otimes
		\underbrace{\tau_3 \otimes \cdots \otimes \tau_3}_{D-1} \\
	\Gamma_{D+2} & = & \mbbf{1} \otimes \tau_2 \otimes
		\underbrace{\tau_3 \otimes \cdots \otimes \tau_3}_{D-2} \\
	\vdots       &   & \\
	\Gamma_{D+i} & = &
		\underbrace{\mbbf{1} \otimes \cdots \otimes \mbbf{1}}_{i-1} \otimes
		\tau_2 \otimes \underbrace{\tau_3 \otimes \cdots \otimes \tau_3}_{D-i}
		\\
	\vdots       &   & \\
	\Gamma_{2D}  & = &
		\underbrace{\mbbf{1} \otimes \cdots \otimes \mbbf{1}}_{D-1}
		\otimes \tau_2
	\end{array}
\end{eqnarray}
where $i$ runs from $1$ to $D$.
It can be easily checked that $\Gamma_n$'s satisfy the relation
\begin{eqnarray}
	\left\{ \Gamma_n, \Gamma_m \right\} = 2 \delta_{nm}
	\qquad (n, m = 1, \ldots, 2D).
\end{eqnarray}
Moreover, we can see that the matrix $\Gamma_i$ is real
and the matrix $\Gamma_{D+i}$ is pure imaginary:
\begin{eqnarray}
	\Gamma_i^\ast     = \Gamma_i, \qquad
	\Gamma_{D+i}^\ast = - \Gamma_{D+i} \qquad (i = 1, \ldots, D).
\end{eqnarray}
From the anti-commutation relation, we find the hermite matrix
$\Gamma_{2D+1}$ which anti-commutes with all $\Gamma_n$'s:
\begin{eqnarray}
	\Gamma_{2D+1} & = & (-i)^D \Gamma_1 \Gamma_2 \cdots \Gamma_{2D}
						\nonumber \\ 
	              & = & \tau_3 \otimes \tau_3 \otimes \cdots \otimes \tau_3.
\end{eqnarray}
Clearly, the matrix $\Gamma_{2D+1}$ is hermitian and the square of this matrix 
is equal to the unit matrix. Thus, we have
\begin{eqnarray}
	\begin{array}{rcl}
		\Gamma_{2D+1} \Gamma_i^\ast \Gamma_{2D+1}     & = & - \Gamma_i, \\
		\Gamma_{2D+1} \Gamma_{D+i}^\ast \Gamma_{2D+1} & = & \Gamma_{D+i}.
	\end{array}
\end{eqnarray}
In $(1+D)$ dimensions, the relation Eq.(\ref{eq:XY_1}) is rewritten as follows:
\begin{eqnarray}
	\left\{
	\begin{array}{rcl}
		X_i^\ast \Gamma & = & - \Gamma X_i, \\
		Y_i^\ast \Gamma & = & \Gamma Y_i.
	\end{array}
	\right.
	\label{eq:XY_i}
\end{eqnarray}

\begin{figure}[htbp]
    \begin{center}
        \epsfig{file=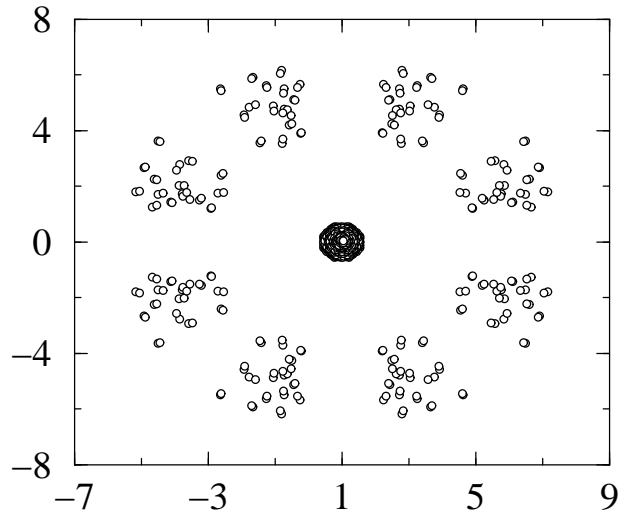}
    \end{center}
    \caption{The spectrum of $\Lambda(x)$ in the complex plane
        on a $4 \times 4 \times 4$ lattice in SU(2) gauge group.}
    \label{fig01}
\end{figure}

\noindent
Then, for the representation of Eq.(\ref{eq:XYGamma2}),
\begin{eqnarray}
	X_i = \sqrt{\frac{1}{r_E}} \Gamma_i, \qquad
	Y_i = \sqrt{\frac{1}{r_E}} \Gamma_{D+i} + \Gamma_{2D+1},
\end{eqnarray}
we find
\begin{eqnarray}
	\Gamma = \Gamma_{2D+1}.
\end{eqnarray}
Since the eigenvalues of $\Lambda(x)$ always consist of complex conjugate pairs
and degenerated ones on real axis, we conclude the determinant
of our fermion matrix is positive in SU(2) gauge fields in any dimensions.

Now we show a numerical evidence.
Fig.\ref{fig01} shows the spectrum of our fermion matrix
in a typical background configuration of link variables for SU(2) gauge group
in $(1+2)$ dimensions.
We find that
their distribution is symmetric with respect to the real axis as expected.
Similarly in $(1+3)$ dimensions we can numerically confirm the symmetry with
respect to the real axis, and the positivity of the determinant.

%%%%%%% discussion and summary %%%%%%%
\section{discussion and summary}

In the previous paper \cite{Our-Simulation1} we reported analytical
and numerical results on the fermion determinant of our new action
in $(1+1)$ dimensions.
In the case of U(1) gauge group,
we were faced with the problem of convergence in numerical simulations.
The cause of the poorness of the convergence is that the summation
of the $\det (1 - S_0^\dagger(x) U_E(x))
= \det (1 - e^{i\Theta} \tilde{S}_0^\dagger(x) U_E(x))$
over arbitrary phase angle $\Theta$ is canceled out accidentally.
The element $e^{i \Theta}$ comes from
the one sided time difference operator $S_0^\dagger(x)$
with $\theta_0(x)$ replaced by $\theta_0(x) + \Theta$,
i.e. $S_0^\dagger(x) = e^{i\Theta} \tilde{S}_0^\dagger(x)$,
where $\theta_0(x)$ is defined by $U_{x, x+\hat{0}} = e^{i \theta_0(x)}$
\cite{Our-Simulation1}.
Therefore we must have control of the phase angle $\Theta$
in order to get good convergence in the $(1+1)$-dimensional U(1) gauge theory.
In fact we analytically showed that
our fermion determinant is real for all configurations
and positive for most configurations under the T-condition
($\theta_0(x) = 0$), which corresponds to the temporal gauge condition
on the infinite lattice, or the GT-condition
($\sum_x \theta_0(x) = n\pi$ : $n = \mbox{even}$), which is achieved by
a gauge transformation on the infinite lattice.
It was also verified numerically.
On the other hand we got good convergence without any conditions
in $(1+1)$-dimensional SU(N) case, because the element like $e^{i \Theta}$
does not belong to the SU(N) group.

\begin{figure}[htbp]
    \begin{center}
        \epsfig{file=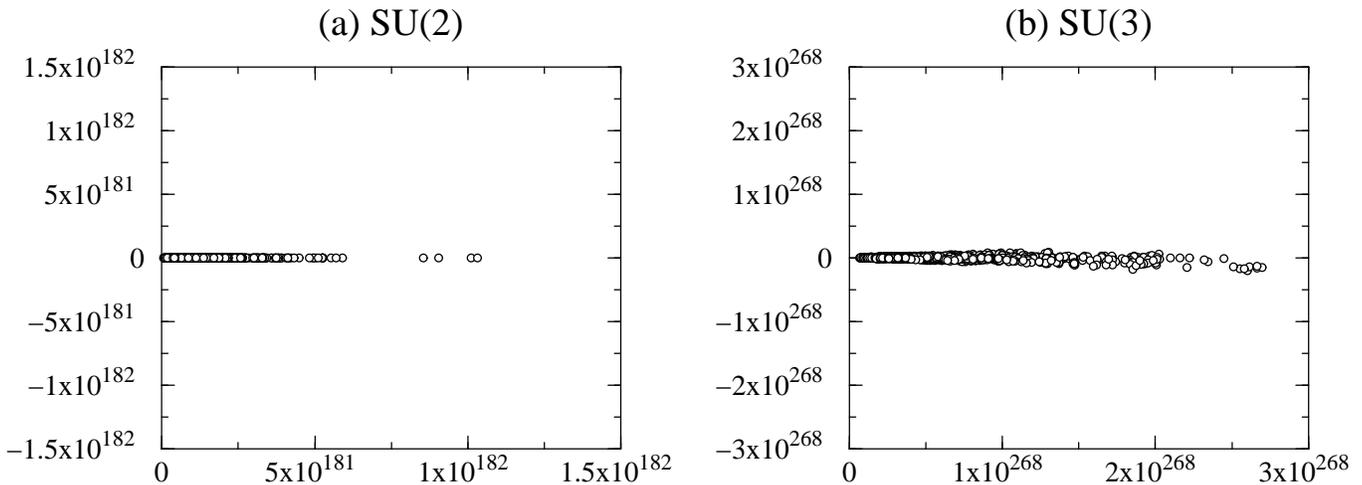}
    \end{center}
    \caption{The distribution in the complex plane of our fermion determinants
        in a) SU(2) and b) SU(3) for each configurations
        of 2000 Monte Carlo iterations after getting good equilibrium,
        i.e. after 2000 iterations, on a $4 \times 4 \times 4$ lattice
        at $\beta = 2.0$.}
    \label{fig02}
\end{figure}

The above discussion is applicable to higher dimensions to a certain extent.
In SU(2) group our fermion determinant is analytically shown
real and positive in any dimensions.
In Fig.\ref{fig02}(a) we give the numerical evidence
that the determinant is real and positive in $(1+2)$ dimensions.
In the case of SU(3) group, we cannot prove the reality of the determinant.
But from Fig.\ref{fig02}(b) we see
that the distribution of the determinant is concentrated near the real axis
without any conditions and the phase angle of the determinant is small.
We have obtained similar results in $(1+3)$ dimensions.
When the phase angle of the fermion determinant is small enough,
we can neglect the phase factor
and make use of $| \det \Lambda(x) |$ instead of $\det \Lambda(x)$.
In the above numerical simulations link variables are updated by
the Metropolis method and determinants are calculated by the LU decomposition.
So there are no systematic errors in the determinants.

In conclusion, we believe that our new fermionic action
is a profitable formulation for the numerical simulations
of SU(2) and SU(3) lattice gauge theory.

%%%%%%% references %%%%%%%

\end{document}